%% file: main.tex
\documentclass[copyright,creativecommons]{eptcs}
\usepackage{breakurl}
\usepackage{graphicx}

\usepackage{txfonts}
\usepackage{pxfonts}

\input{draw}

\title{Graph Creation, Visualisation and Transformation}
\author{Maribel Fern\'andez and Olivier Namet
\institute{King\'s College London\\
 Department of Computer Science\\
 Strand, London WC2R 2LS, U.K.}
\email{olivier.namet@kcl.ac.uk}
}
\def\titlerunning{Graph creation, visualisation and transformation}
\def\authorrunning{Fern\'andez and Namet}
\begin{document}
\maketitle

\begin{abstract}
  We describe a tool to create, edit, visualise and
  compute with interaction nets --- a form of graph rewriting
  systems. The editor, called GraphPaper, allows users to create and
  edit graphs and their transformation rules using an intuitive user
  interface. The editor uses the functionalities of the TULIP system,
  which gives us access to a wealth of visualisation
  algorithms. Interaction nets are not only a formalism for the
  specification of graphs, but also a rewrite-based computation
  model. We discuss graph rewriting strategies and a language to express
  them in order to perform strategic interaction net rewriting.
\end{abstract}

\section{Introduction}

Graph representation and graph transformations are important in
Computer Science. It is well known that graphical formalisms have
clear advantages as modelling tools, in particular in the earlier
phases of system specification and development. Graphical formalisms
are more intuitive and make it easier to visualise the system, whether
in theoretical or practical domains. For example, consider the textual
representation of proofs in the sequent calculus~\cite{GentzenG:invld}
versus proof nets~\cite{GirardJY:linl}, the entity-relationship
diagrams~\cite{BarkerR:case} used to specify a relational database versus the tables, etc.

On the negative side, there are some well-known implementation
problems when dealing with graphical formalisms (pattern-matching is
not an easy problem, see for instance~\cite{UllmanJ:subgip}), and
graph rewriting can be inefficient in general.  Graphical editors and
graphical programming environments exist, but none of the available
tools to manipulate graphs provides a unified framework to create,
edit and visualise graphs, and to define dynamic transformations.
Ideally, such a tool should allow users to create graphs, edit them
and define different views, export to different formats (e.g., image
files, encapsulated postscript, latex macros, etc.), and should also
be able to model some notion of computation, allowing the user to
evaluate parts of their graphs using transformation rules plugged into
the tool.

In this paper, we describe the design of an editor to draw graphs and
their transformation rules, and its integration into a tool that can
be used to visualise graphs and their associated computations. The
editor, called GraphPaper, is tailored for the design of interaction
net~\cite{LafontY:intn} systems. These are graph rewriting systems
where nodes are connected by edges attached to ports, and the
rewriting rules are constrained in order to ensure useful rewriting
properties, such as confluence, by construction.  The user interface
of GraphPaper mimics the operations that users perform to draw graphs
manually in paper. Several checks are automatically performed to
facilitate the construction of interaction nets and rewriting rules.

Graph rewriting needs complex pattern matching, but in the case of
interaction nets, the pattern-matching algorithm is simpler due to the
restricted form of the left-hand sides of the rules. Still, there are
several problems that have to be solved in a graphical editor, due to
the changes in the layout that may arise after a rewriting step.
Furthermore, users of graph rewriting systems often need to define
specific strategies dictating how and when rules are applied, thus a
formal language is needed to express these strategies.

The GraphPaper editor simplifies the task of creating interaction nets
and rules, and provides an interface with the
TULIP\footnote{http://www.tulip-software.org} system for analysis and
visualisation. Strategic interaction net rewriting, tracing and
debugging are implemented as extensions of the TULIP system.

TULIP is an environment for graph visualisation which provides us with
algorithms to display graphs in various formats, and to check static
properties of graphs. We use GraphPaper to create and edit our graph
systems, which are then exported to TULIP.  Although TULIP provides
powerful visualisation and analysis algorithms, the standard version
of TULIP has no features relating to graph rewriting and the notion of
port, which is essential to our rewriting systems, is not
available. However, TULIP is extensible and plugins can be added to
implement graphs with ports, such as interaction nets, 
as well as pattern matching and  rewriting. 
We describe the main components of the system below.

In future work, we will also include a mechanism for tracing graphs
during the rewriting process to allow greater control and facilitate
the debugging of the rewriting system.  This is work in progress
within the PORGY collaboration between INRIA Bordeaux and King's
College London, which focuses on graph visualisation and rewriting
strategies for interaction nets and port graphs~\cite{helene} in
general.

\paragraph{Related Work}

There are many graphical editors available, but only a few of them
allow the user to specify dynamic information in the form of 
graph rewriting rules. Below we discuss four systems
that include this functionality and are directly related to our work.

The Interaction Net Laboratory
(INL)\footnote{http://inl.sourceforge.net/} developed by De Falco is a
graphical editor for interaction nets. It has a rich feature set with
a Net editor and a Rule editor.  Cells can be created and have
properties such as ``title'' and ``colour''.  Feature wise, INL is
quite complete for creating and editing nets and rules but the
user-interface has some limitations.  Clicking on objects does not
result in immediate visual feedback.
To add cells to the net the user  has to
click on the cell in the list to the left and then click somewhere on
the net. To add more than one of the same kind of cell,
one has to keep on re-clicking on the cell that is in the list. 

The Graphical Interpreter for Interaction Nets developed by
Lippi~\cite{LippiS:in2} imports a text based
representation of a net and a set of reduction rules and creates a
graphical net.  The user can then choose to reduce the net step by
step or to perform all possible reductions in one go. While this
program is useful for displaying a net and visualising its
reduced form, the user still needs to input and edit the nets using 
a text-based language. 

INblobs\footnote{http://haskell.di.uminho.pt/jmvilaca/INblobs/}  developed by
Vila\c{c}a et al.~\cite{VilacaM:RULE07}  defines itself as \emph{an editor
 and interpreter for Interaction Nets}.  Feature wise, INblobs is on
par with INL but suffers from similar problems.

 PROGRES is a programming environment based on graph grammars
  developed at the University of Technology Aachen\footnote{See
  \texttt{http://www-i3.informatik.rwth-aachen.de/tikiwiki/tiki-index.php}}. It
  is built around an executable specification language based on a
  specific kind of graph rewriting rules. The
  environment provides a graphical editor for the specification
  language and a translator into C
  and Tcl/Tk-code.  The tool seems to focus on the specification
  language, which is expressive enough to allow the user to model
  complex systems using graphs. Our goals are different: we focus on
  the graphical editor and in the principles behind the design of a
  graphical interface for the representation of graphs.

\paragraph{Overview of the paper} 
Section~\ref{sec:prelim} provides a concise introduction to
interaction nets and rewriting strategies.
In Section~\ref{sec:grap} we describe  GraphPaper's novel ``digital paper'' user
interface. Section~\ref{sec:grar} deals with visualisation and
rewriting; we describe the architecture of the system and its
implementation via TULIP. Section~\ref{sec:conc} contains conclusions
and directions for future work.

\section{Preliminaries}
\label{sec:prelim}

\subsection{Interaction Nets}

A system of interaction nets is specified by a set $\Sigma$ of symbols
with fixed arities, and a set $\IR$ of interaction rules.  An
occurrence of a symbol $\alpha\in\Sigma$ is called an \emph{agent}. If
the arity of $\alpha$ is $n$, then the agent has $n+1$ \emph{ports}: a
\emph{principal port} depicted by an arrow, and $n$ \emph{auxiliary
  ports}. Such an agent will be drawn in the following way:

\begin{net}{40}{40}
\putalpha{20}{20}
\putDvector{20}{10}{10}
\putline{12.6}{27.4}{-1}{1}{10}
\putline{27.4}{27.4}{1}{1}{10}
\puttext{20}{35}{$\cdots$}
\put(2.6,38){\makebox(0,0)[br]{$x_1$}}
\put(37.4,38){\makebox(0,0)[bl]{$x_n$}}
\end{net}

Intuitively, a net $N$ is a graph (not necessarily connected) with
agents at the vertices and each edge connecting at most 2 ports. The
ports that are not connected to another agent are \emph{free}. There
are two special instances of a net: \label{wiring}a wiring (no agents) and the empty
net; the extremes of wirings are also called free ports. The
\emph{interface} of a net is its set of free ports.

An interaction rule $((\alpha,\beta) \Lra N) \in \IR$ replaces a pair
of agents $(\alpha,\beta)\in \Sigma\times\Sigma$ connected together on
their principal ports (an \emph{active pair} or \emph{redex}, written
$\alpha \bowtie \beta$) by a net $N$ with the same interface. Rules
must satisfy two conditions: all free ports are preserved during
reduction (reduction is local, \ie only the part of the net
corresponding to the redex is modified, no global modifications are
required), and there is at most one rule for each pair of
agents. Because of this last restriction, a rule is fully determined
by its left hand-side; such a rule will thus be sometimes denoted by
$\alpha \bowtie \beta$ as well. The following diagram shows the format
of interaction rules ($N$ can be any net built from $\Sigma$).

\begin{net}{200}{60}
\putalpha{20}{20}
\putbeta{60}{20} \putRvector{30}{20}{10}
\putLvector{50}{20}{10} \putline{12.6}{27.4}{-1}{1}{10}
\putline{12.6}{12.6}{-1}{-1}{10} \putline{67.4}{27.4}{1}{1}{10}
\putline{67.4}{12.6}{1}{-1}{10} \puttext{5}{23}{$\vdots$}
\puttext{75}{23}{$\vdots$} \put(0,0){\makebox(0,0)[br]{$x_1$}}
\put(0,40){\makebox(0,0)[tr]{$x_n$}}
\put(80,0){\makebox(0,0)[bl]{$y_m$}}
\put(80,40){\makebox(0,0)[tl]{$y_1$}} \puttext{102}{20}{$\Lra$}
\putbox{140}{0}{50}{40}{$N$} \putHline{130}{10}{10}
\putHline{130}{30}{10} \puttext{135}{23}{$\vdots$}
\putHline{190}{10}{10} \putHline{190}{30}{10}
\puttext{195}{23}{$\vdots$} \put(125,5){\makebox(0,0)[br]{$x_1$}}
\put(125,35){\makebox(0,0)[tr]{$x_n$}}
\put(205,5){\makebox(0,0)[bl]{$y_m$}}
\put(205,35){\makebox(0,0)[tl]{$y_1$}}
\end{net}

We use the notation $\Lra$ for the one-step reduction relation, or
$\Lrawith{\alpha \bowtie \beta}$ if we want to be explicit about the
rule used, and $\Lra^*$ for its transitive and reflexive closure. If a
net does not contain any active pairs then we say that it is in normal
form. The key property of interaction nets, besides locality of
reduction, is that reduction is strongly confluent. Indeed, all
reduction sequences are permutation equivalent and standard results
from rewriting theory tell us that weak and strong normalisation
coincide (if one reduction sequence terminates, then all reduction
sequences terminate).
We refer the reader to~\cite{LafontY:intn} for
more details and examples.

\subsection{Rewriting Strategies}
\label{strat}

A graph rewriting system may have a potentially large set of rules to
apply to a graph.  The order in which rules are applied can
greatly alter the end graph when general graph rewriting is
considered. In the case of interaction nets, the strong confluence
property ensures that all reduction sequences to full normal form are
equivalent.  However, this is not the case if we use a notion of
reduction that does not reach a full normal form (for instance,
reduction to interface normal form~\cite{FernandezM:calin}).  Also,
for interaction nets, even if the end graph does not change, the size
and layout of the graph during the rewriting process can differ
depending on what rules and where they are applied first. Users may
therefore want to not just blindly apply rules but to create a
strategy around these rules to direct the rewriting.

Strategic rewriting has been studied for term rewriting systems, and
there are languages that allow the user to specify a strategy and to
apply it~\cite{elan1, Vis01.rta}.  In this paper we will define a
language to define strategies for graph rewriting systems, where not
only the strategy needs to take into account rules and sequences of
rules but also location and propagation in a graph (the latter is
complicated by the fact that in a graph there is no notion of a root,
so standard term rewriting strategies based on top-down or bottom-up
traversals do not make sense in this setting). Because of this, we
develop a specific language to deal with strategies for interaction
nets, which can be also applied to general graph rewriting systems.



\section{GraphPaper}
\label{sec:grap}
\subsection{Digital Paper}
\label{digitalpaper}

In the past few years, new technology has shown that new Human Interface Devices (HID) can be an efficient way of dealing with digital information. There is the example of video games with Nintendo's Wii and DS consoles. One allows users to perform natural motions to convey commands (swinging your arm to swing a sword) while the other lets the user draw to the screen using a tablet. These methods of interaction have been hugely popular across all sorts of demographics.
Also many new mobile devices now come with touch-screen and therefore software makers have had to rethink the way users interact with their phones. Much like the Nintendo consoles, touch screen phones (like the Apple iPhone) have also been very popular across all demographics.

It is indeed more natural to push and pinch a map around on a touch screen to move it than it is using a mouse and icons. This is because these new HID mimic natural motions.
The data that are represented in this case are graphs where the natural instinct is to use a pen and paper to draw them. We will take this intuition and try to apply it to the HCI ideology of GraphPaper. A user should feel like (s)he is using a pen and paper but with the dynamic advantages of a computer.

For graphs, agents and edges are the two types of data that the user will
interact with. This allows the tool to have a simple interface. There
is no need to have a toolbar to activate whether the user is trying to
interact with agents or edges; it is possible for the tool to
deduce what the user is interacting with based on the context of the
interaction. For example, edges need to be created between ports so
cannot be created independently. This means that if the user is trying
to work on an empty part of the canvas, they are trying to create
agents or a wiring.
When the
user tries to interact with ports of an agent, then we know that the
goal is to create or edit edges to and from that agent.
For more complex operations on graphs shape recognition is needed (see, for instance,~\cite{HammondT:LADDER}). Once a shape is recognised, the user's intention can be deduced by the context and location where the shape was drawn. 

Here is an outline of shapes and some of their actions depending on context:
\begin{itemize}
\item
Circular: In GraphPaper, circular shapes are used to represent nodes in the graph,
but may also have other meanings, depending on the state of the canvas
where the shape is being created. We show below two examples.

\begin{center}
\includegraphics[width=150px]{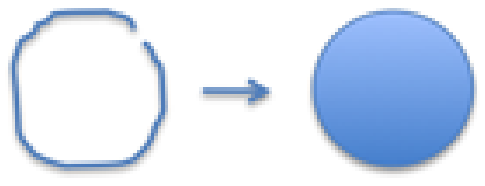}
\\\emph{If created on an empty space then a new agent is created there.}
\end{center}

\begin{center}
\includegraphics[width=200px]{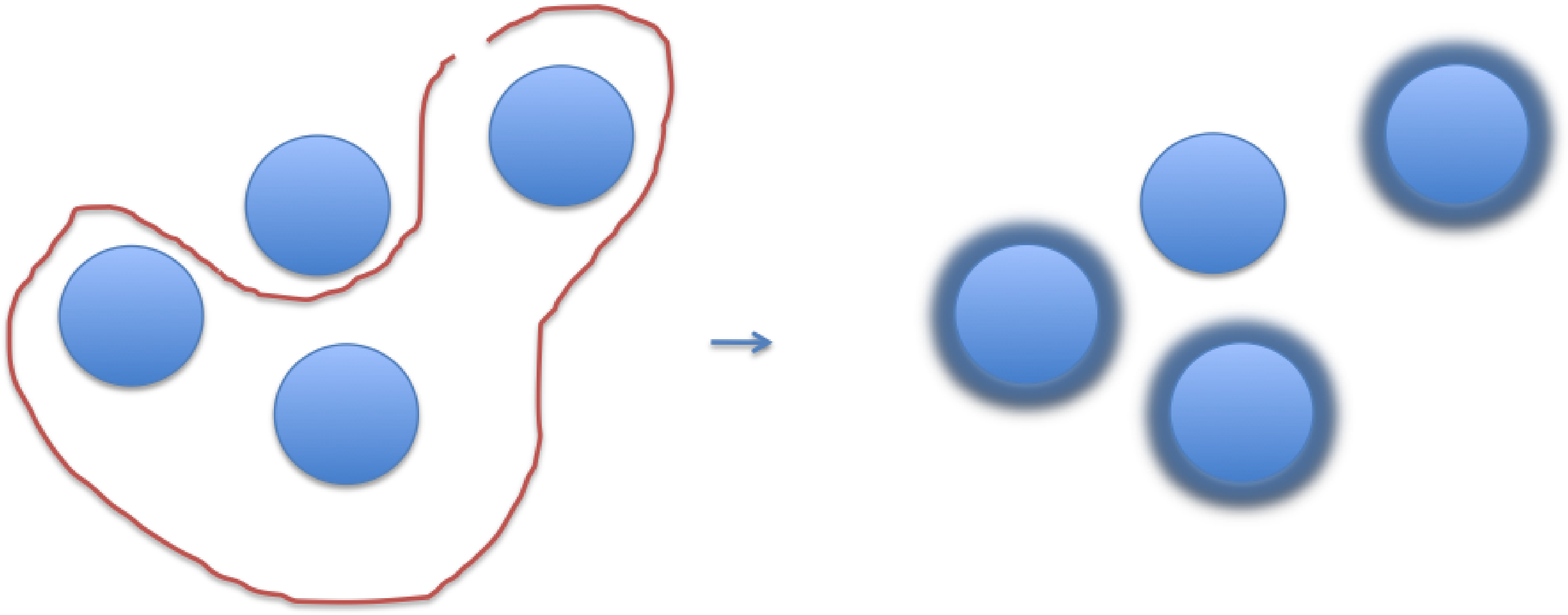}
\\\emph{If other agents are inside the bounds of the circle, the user's intention was to select those agents. The circle becomes a lasso selection tool.}
\end{center}

\end{itemize}

\begin{itemize}
\item
Line: In GraphPaper, lines are interpreted in different ways depending
on the context. We give two examples below.

\begin{center}
\includegraphics[width=200px]{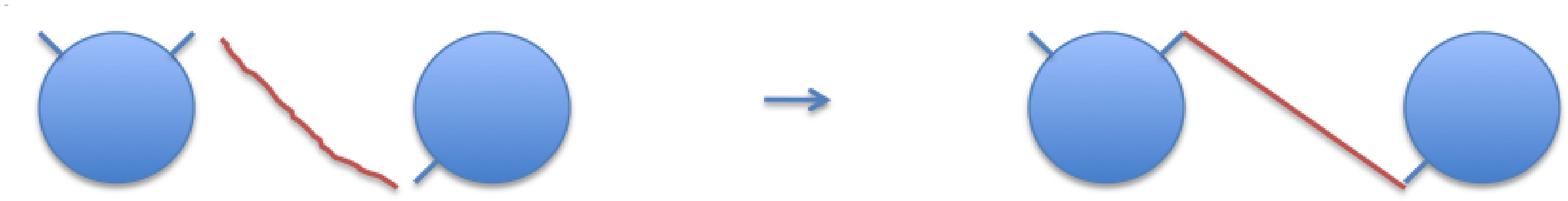} \\
\emph{If a line is drawn between two
  ports, an edge is created between them}
\end{center}

\begin{center}
\includegraphics[width=150px]{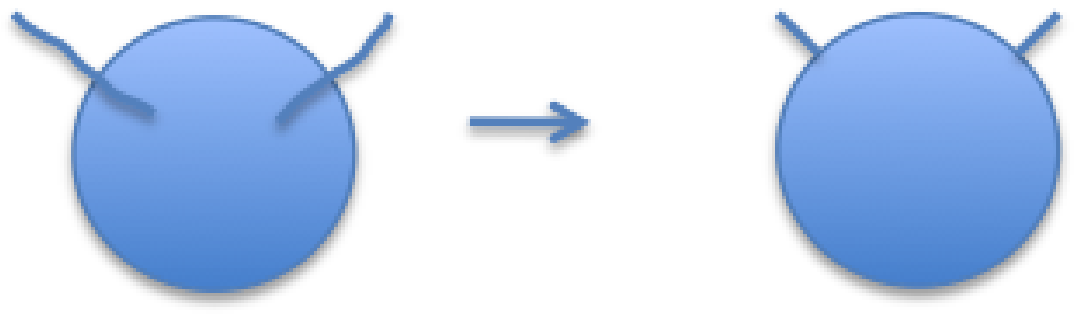}
\\\emph{If a line is drawn from inside an agent to just outside of it, then a port is created at that location on the agent.}
\end{center}
\end{itemize}

In the case of interaction nets, all agents must have one principal
port and zero or more auxiliary ports. In this case, when agents are
created, a principal port is automatically added. 

\begin{figure}
\begin{center}
\includegraphics[width=100px]{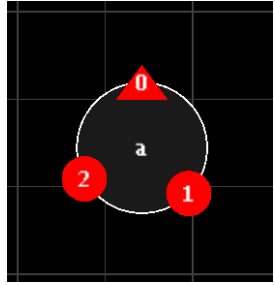}
\caption{A node representing an occurrence of the agent $a$, with 
two auxiliary ports.} 
\label{fig:agentports}
\end{center}
\end{figure}

Users can label nodes with names simply by writing the name inside the
circle that represents the agent. When another node with the same name
is created, GraphPaper automatically adds the corresponding number of
ports. Ports are uniquely numbered with port 0 representing the principal port. To easily distinguish this port visually from the auxiliary ports, GraphPaper draws it as a triangle, as depicted in Figure~\ref{fig:agentports}.

GraphPaper has two \emph{states} that the user can be in when it comes to the creation of graphs. A \emph{view/edit} state and a \emph{draw} state:
\begin{itemize}
\item
The \emph{view/edit} state will allow the user to modify already created agents and edges. For example, moving agents around, renaming agents and ports, etc. The user will also be able to modify the view by panning around and zooming in and out.
\item
The \emph{draw} state is where the user can create and delete new objects by drawing to the screen. See the previous examples of creation.
\end{itemize}



\subsection{Drawing Rules}

Graph rewriting rules consist of a left hand side graph and a right hand
side graph, and a mapping that defines the relation between the interfaces
of the graphs. In the case of interaction nets, both sides must have
the same interface, and the left hand side must be a graph consisting
of two agents connected through their principal ports.  To draw
an interaction rule using GraphPaper, the user simply draws the two
agents in the left hand side in the standard way, and then lasso
selects them. In this way GraphPaper moves to the rule drawing mode,
and the user can continue drawing the right hand side. The
correspondence between ports of the interface in the left hand side
and in the right hand side is explicitely indicated by joining the
ports (GraphPaper will show these links in a different colour).
GraphPaper will then deduce this is a rule, and will isolate it to a
part of the \emph{paper} where users can still have access to it if
they need to modify it. 

\begin{figure}
 \begin{center}
\includegraphics[width=200px]{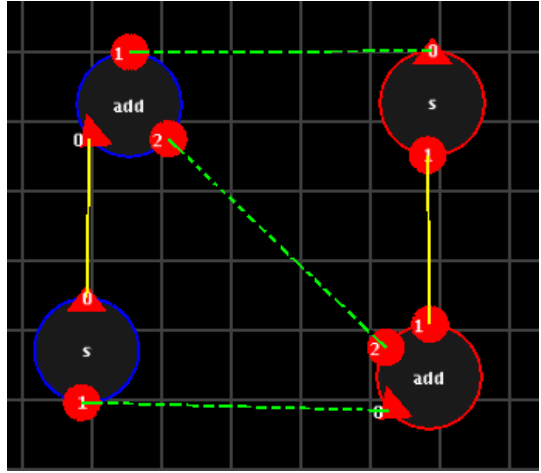}
\end{center}
\caption{An interaction rule between the $S$ and $add$ agents in GraphPaper}
\label{fig:rule}
\end{figure}

Figure~\ref{fig:rule} shows an example screen of a rule. 
Green dotted wires represent the correspondences between interface ports in the  
left-hand side and right-hand side nets.

\section{Graph Rewriting}
\label{sec:grar}
For the visualisation and transformation of graphs, GraphPaper uses
the functionalities available through the
TULIP\footnote{http://www.tulip-software.org} tool developed at INRIA
Bordeaux. We describe below the main features of GraphPaper as a graph rewriting tool.

\begin{itemize}
\item
Pattern Matching: the current version of TULIP does not provide a
pattern-matching algorithm, but it has a built in search function that
can iterate through a graph of potentially very large size. Using this
iteration function, we can then build a
pattern matching algorithm to find possible reductions in a graph. In
the case of general graphs, the high complexity for pattern matching
is greatly aided by the efficient iteration that TULIP provides. For
Interaction Nets, it is simply a case of iterating through all the
edges, finding the ones that connect two principal ports and adding
them to a list. After every reduction, we then just need to iterate
through that new sub-graph and its neighbours to find new active edges
and add those to the list. There is no need to search the entire graph
for new active edges since the reduction only affects the new
sub-graph and its neighbours.
\item
TULIP is capable of displaying graphs with over 1,000,000 elements and
can display graphs of that size in real time. GraphPaper inherits this power,
allowing the user to work with graph rewriting systems where graphs
grow after each rewrite step. 
\item
Visualisation: TULIP provides dynamic visualisation of graphs, that is to say if
a sub-graph is added or removed, the remaining graph dynamically
changes its appearance to a chosen visual model (for example a cone
tree, circular, planar). This is particularly useful in the case of
rewriting graphs since once a rewrite occurs, the graph might need to
be redrawn to accommodate for extra or lack of spacing.

\item Strategies: As discussed in Section~\ref{strat}, strategies for
  graph rewriting need to specify the way rules will be applied and
  must also be aware of \emph{location}. We propose to use expressions
  generated by the following grammar:
$$S := id \mid fail \mid R \hspace{5pt}|\hspace{5pt} S;S \hspace{5pt}|\hspace{5pt} S\parallel{}S \hspace{5pt}|\hspace{5pt} S^* \hspace{5pt}|\hspace{5pt} S~ or~ S$$ 

where $id$ is the identity (which never fails and always leaves the
graph unchanged); $fail$ is a strategy that always fails (it leaves
the graph unchanged and returns failure); $R$ is an expression of the
form $R_i(subgraph, depth)$ that denotes the application of rule $R_i$
in the graph $subgraph$ (which can be selected using the graphical
interface) or its neighbours up to the given \emph{depth} (as
explained below), this strategy may fail is the rule is not applicable
in the designated subgraph; $S_1;S_2$ represents sequential
application: apply $S_1$ and if successful then apply $S_2$ (if either
of them was unsuccessful the result is a failure); $S_1\parallel{}S_2$
represents simultaneous application (both strategies must be applied
at the same time); $S^*$ means ``apply S as many times as possible in
a row''; and $S_{1}~ or ~ S_2$ means apply $S_1$, if it fails then
apply $S_2$ (not both).  Note that when defining a graph rewriting
strategy as a parallel composition (i.e., simultaneous application of
two strategies), conflicts may arise. However, in the case of
interaction nets, the constraints on interaction rules imply that each pair of
agents can only be involved in one interaction rule, hence all redexes can
be simultaneously reduced without conflict.

The basic blocks to build a strategy are a rule and the identity. When
we indicate that a rule will be applied, we must also provide the
location where the rule should be applied. This is given by the
arguments $subgraph$ and $depth$, i.e., we specify a subgraph, and the
$depth$ represents how far one should look through the $subgraph$'s
neighbours for a possible application of the rule. If we want it to be
strict, i.e. apply the rule in the given subgraph only, we use $0$. To
look as far as possible starting from the given subgraph we use the
value $-1$. The expression $R_i(subgr,1)$ indicates that we want to
apply the rule $R_i$ in the subgraph that includes $subgr$ and all the
neighbours at distance 1. To look $n$ steps out of the $subgraph$ then
set $depth$ to $n$. In the case of interaction nets, the $depth$
search only follows principal ports of the nodes.

For convenience, when composing strategies we allow the user to factor
out the common sub-expressions: If we compose strategies that have the
same location (same subgraph and depth) we can write these parameters
only once (e.g. $(R_1;R_2)(subgr,0)$ indicates that we want to apply
$R_1$ and then $R_2$ to the subgraph $subgr$), and if we wish to apply
the same strategy at several locations in the graph, we can write for
example $(R_1;R_2)[(subgr_1,0),(subgr_2,0)]$, meaning that we need to
apply $R_1$ followed by $R_2$ in $subgr_1$ and also in $subgr_2$.

We also define auxiliary functions $Interface$($subgraph$) which
returns a graph containing the interface nodes of $subgraph$, and
$Successors(subgraph)$. The first, used with $R_i()$, allows the user
to easily write strategies that give priority to rewriting steps at
the interface of the $subgraph$: for example,
$R_1(Interface(subgraph),1)$ tries to apply $R_1$ on nodes of the
interface and their neighbours.  This is useful when computing
interface normal forms of interaction nets.

As an example the following strategy $(R_1 or
R_2);R_3^*[Interface(sub_{1}),0]$ is a strategy that will take the
strict interface of the subgraph $sub_{1}$ and either apply $R_{1}$ or
$R_{2}$ once and then apply $R_{3}$ as many times as possible.

\item Trace: Each time a rewrite is performed, a new graph is
  created. To keep track of the rewriting history, we use a Trace that
  will store all the different graphs, and if one graph is the result
  of a rewrite of another then an edge is created from the latter to
  the former with the rule and location of the rewriting as its
  label. Since more than one rewrite is possible at any one time, the
  Trace will branch for each one, allowing the user to see all the
  possibilities. The Trace will therefore take on the shape of a tree,
  as depicted in Figure~\ref{fig:trace}, where the main components of the system are shown.

\begin{figure}
 \begin{center}
\includegraphics[width=350px]{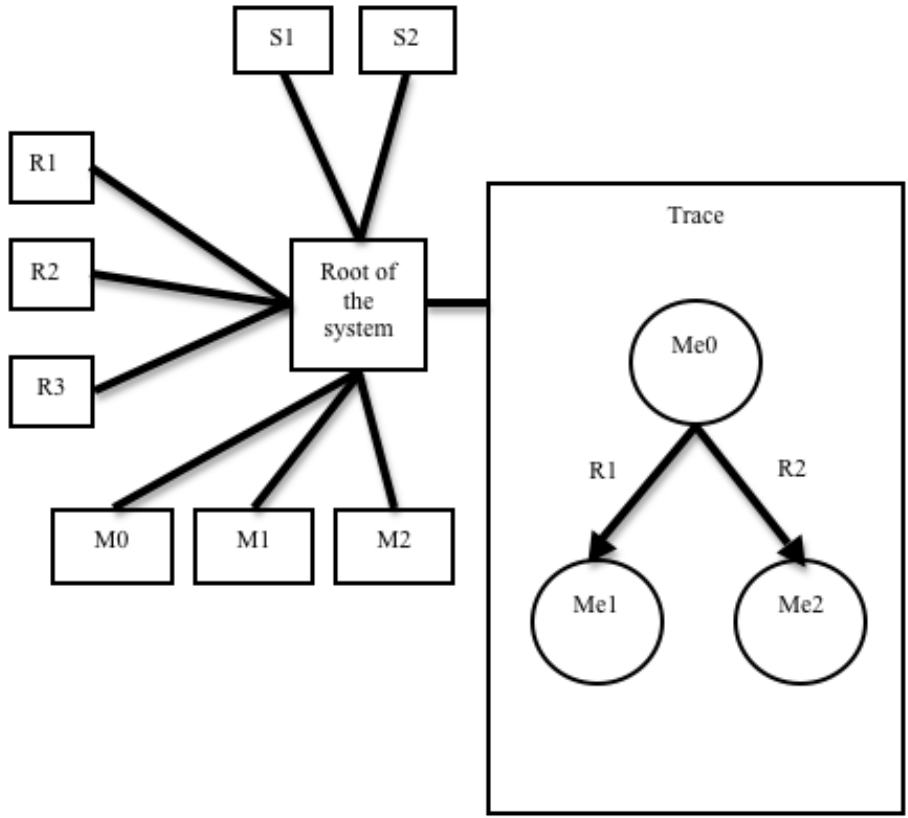}
\end{center}
\caption{A schematic description of the main components of the system.}
\label{fig1}
\label{fig:trace}
\end{figure}

\item Architecture:
 Since TULIP is very efficient when it comes to storing graphs, we define and represent everything using a main graph which we call root graph, see Figure~\ref{fig1}.  The set of rules ($R_1,R_2,\ldots$) and strategies ($S_1, S_2,\ldots$) are also stored as subgraphs of the root graph and each have a unique name. A base model M0 is also created as a subgraph of the root graph and holds the initial state of the graph the user will be rewriting on. The trace is also subgraph  of the root. 
Below we show a GUI concept for TULIP perspective with an example trace:
 \begin{center}
\includegraphics[width=400px]{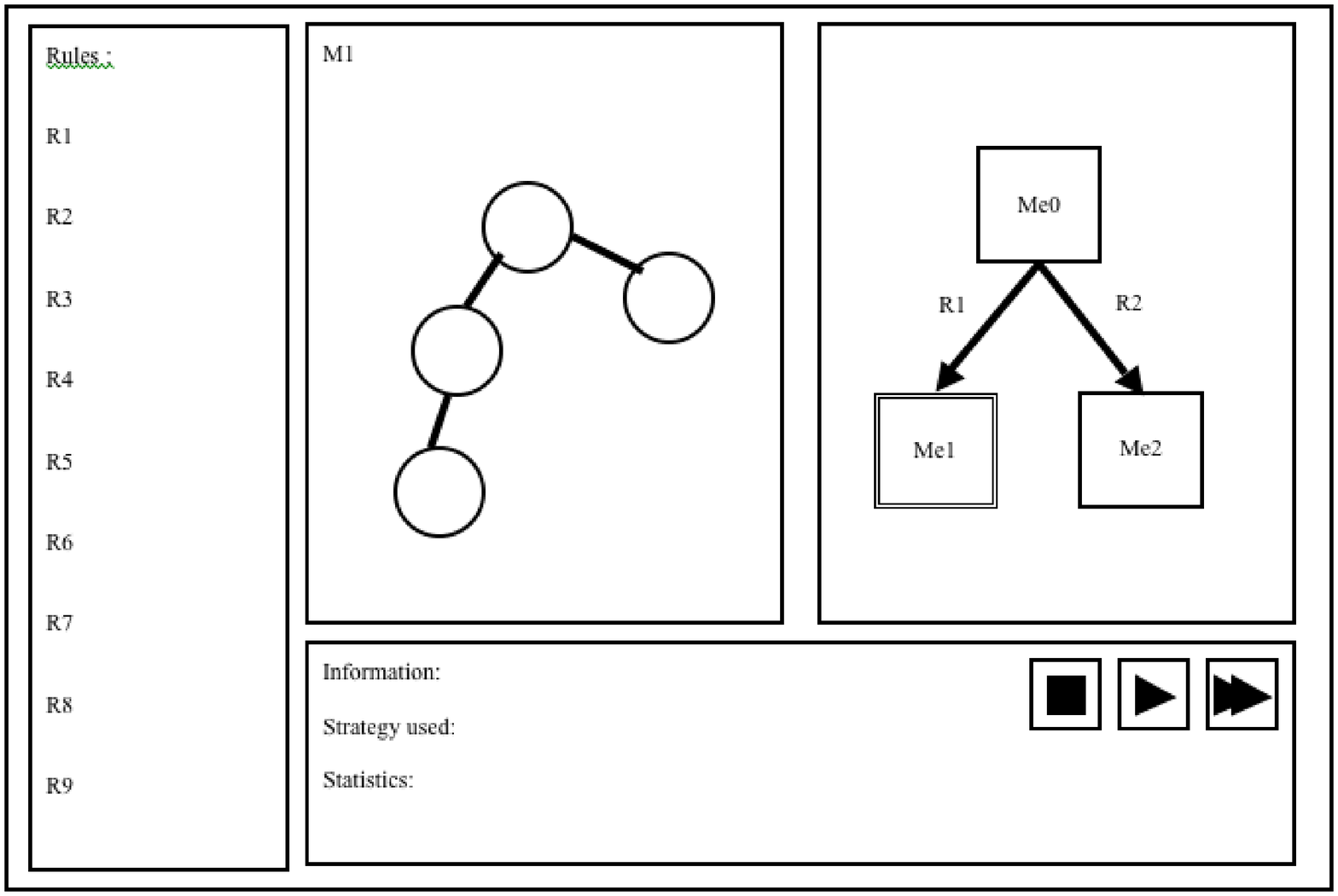}
\label{fig2}
\end{center}
In this particular example, a graph $M0$ was created by the user (stored as $Me0$ in the trace) and  rules $R1$ and $R2$ were applied to $M0$. The user then selected $Me1$ in the $Trace$ to get a closer look at $M1$.
\end{itemize}

\section{Conclusion and Future Work}
\label{sec:conc}
GraphPaper functions as a stand-alone tool used specifically to create and edit graphs and their rules and strategies. The graph system created can then be exported into TULIP where the rewriting will happen based on a selected strategy. The tool will generate a $Trace$ of the rewriting as it happens and then allow the user to observe any point of the rewrite in more detail.
The ease of use of GraphPaper combined with the power of TULIP and the detailed information provided by the Trace gives the user an environment to work on graph rewriting systems, and interaction nets in particular, efficiently and intuitively. In particular, GraphPaper can serve as an editor for visual programming languages based on interaction nets (see~\cite{HassanA:visin}).

In future and within the PORGY 
collaboration, we hope to develop the tools to support more general forms of 
graph rewriting by implementing a more complex and versatile pattern matching algorithm. We will also extend the interface for rule definition in GraphPaper, in order to represent more general kinds of rules, such as the ones used in PortGraph systems~\cite{helene}.

\bibliographystyle{eptcs} 


\end{document}

%% file: draw.tex
\usepackage{xspace}

\newcommand{\IR}{\mathcal{R}}
\newcommand{\Lra}{\Longrightarrow}
\newcommand{\ie}{\mbox{i.e.}\xspace}

\def\stackrelbis#1#2{\mathrel{\mathop{#2}\limits_{#1}}}
\newcommand{\LLra}{\mathrel=\joinrel\Longrightarrow}
\newcommand{\Lrawith}[1]{\stackrelbis{#1\;}{\LLra}}

\newenvironment{net}[2]{%
\begin{center}%
\begin{picture}(#1,#2)
}{%
\end{picture}%
\end{center}}

\newcommand\putcircle[4]
{\put(#1,#2){\circle{#3}}\put(#1,#2){\makebox(0,0){#4}}}

\newcommand\putbox[5]
{\put(#1,#2){\framebox(#3,#4){#5}}}

\newcommand\putline[5]
{\put(#1,#2){\line(#3,#4){#5}}}

\newcommand\putnline[3]
{\count0 = #2 \advance \count0 by 10%
\put(#1,#2){\line(0,1){#3}}%
\put(#1,\count0){\makebox(0,0){$/$}}}

\newcommand\putvector[5]
{\put(#1,#2){\vector(#3,#4){#5}}}

\newcommand\putaxiom[3]
{\count2 = #3 \count0 = #3 \divide \count0 by 2
\count1 = #1 \advance \count1 by \count0
\put(\count1,#2){\line(-1,0){#3}}%
\put(\count1,#2){\line(0,-1){10}}%
\advance \count1 by -\count2%
\put(\count1,#2){\line(0,-1){10}}%
}

\newcommand\putcut[3]
{\count2 = #3 \count0 = #3 \divide \count0 by 2
\count1 = #1 \advance \count1 by \count0
\put(\count1,#2){\line(-1,0){#3}}%
\put(\count1,#2){\line(0,1){10}}%
\advance \count1 by -\count2%
\put(\count1,#2){\line(0,1){10}}%
}

\newcommand\putHline[3]
{\putline{#1}{#2}{1}{0}{#3}}

\newcommand\putDvector[3]
{\putvector{#1}{#2}{0}{-1}{#3}}

\newcommand\putLvector[3]
{\putvector{#1}{#2}{-1}{0}{#3}}

\newcommand\putRvector[3]
{\putvector{#1}{#2}{1}{0}{#3}}

\newcommand\putbeta[2]
{\putcircle{#1}{#2}{20}{$\beta$}}

\newcommand\putalpha[2]
{\putcircle{#1}{#2}{20}{$\alpha$}}

\newcommand\putdtriangle[3]{\count0=#2 \advance \count0 by 9%
\count1=#1 \advance \count1 by -10%
\count2=#2 \advance \count2 by 15%
\put(#1,#2){\line(-2,3){10}}%
\put(#1,#2){\line(2,3){10}}%
\put(\count1,\count2){\line(1,0){20}}%
\advance \count1 by 4%
\put(\count1,\count2){\line(0,1){10}}%
\advance \count1 by 12%
\put(\count1,\count2){\line(0,1){10}}%
\put(#1,#2){\vector(0,-1){10}}%
\put(#1,\count0){\makebox(0,0){#3}}}

\newcommand\puttext[3]
{\put(#1,#2){\makebox(0,0){#3}}}
